\documentclass[%
 reprint,
 amsmath,amssymb,
 aps,
]{revtex4-1}

\usepackage{graphicx}
\usepackage{dcolumn}
\usepackage{bm}
\usepackage{hyperref}

\usepackage[caption=false]{subfig}
\usepackage{siunitx}
\usepackage{amsthm}
\usepackage{physics}

\DeclareMathOperator{\variance}{Var}

\begin{document}

\preprint{APS/123-QED}

\title{Temporal array with superconducting nanowire single-photon detectors for photon-number-resolution}

\author{Mattias J\"{o}nsson}
\email{matjon4@kth.se}
\affiliation{%
 Department of Physics, KTH Royal Institute of Technology\\
 AlbaNova University Center, SE 106 91 Stockholm, Sweden
}
\author{Marcin Swillo}%
\author{Samuel Gyger}
\author{Val Zwiller}

\author{Gunnar Bj\"{o}rk}%
\email{gbjork@kth.se}
\affiliation{%
 Department of Applied Physics, KTH Royal Institute of Technology\\
 AlbaNova University Center, SE 106 91 Stockholm, Sweden
}%




\date{\today}

\begin{abstract}
We present an experimental realization of a 16 element, temporal-array, photon-number-resolving (PNR) detector, which is a multiplexed single-photon detector that splits an input signal over multiple time-bins, and the time-bins are detected using two superconducting nanowire single-photon detectors (SNSPD). A theoretical investigation of the PNR capabilities of the detector is performed and it is concluded that compared to a single-photon detector, our array detector can resolve one order of magnitude higher mean photon numbers, given the same number of input pulses to measure. This claim is experimentally verified and we show that the detector can accurately predict photon numbers between $10^{-3}$ to $10^{2}$. Our present detector is incapable of single-shot photon-number measurements with high precision since its effective quantum efficiency is $\SI{49}{\percent}$. Using SNSPDs with a higher quantum efficiency the PNR performance will improve, but the photon-number resolution will still be limited by the array size.
\end{abstract}

\maketitle


\section{\label{sec:introduction}INTRODUCTION}

Photon-number-resolving detectors are devices capable of accurately measuring photon numbers in input wave-packets. Having the ability to resolve photon numbers is applicable and highly desirable in imaging applications \cite{Holland1999}, investigation of exceptional points in $\mathcal{PT}$-symmetric systems \cite{QuirozJuarez2019}, quantum key distribution \cite{Cattaneo2018HybridDetectors}, decoy states for quantum key distribution security \cite{Hwang2003QuantumCommunication, Lo2005DecoyDistribution}, measurements in the number basis \cite{Kovalenko2018}, and photon-counting laser-radars \cite{Huang2014}. 


PNR detectors can be divided into two categories, inherent PNR detectors which utilize some process where the quantity used as the output has a well-defined dependence on the incident photon number, and multiplexed PNR detectors which split the input over multiple single-photon detectors. Examples of detectors in the former category are transition edge sensors (TES) \cite{Lita2010SuperconductingDetectors, Fukuda2011, Konno2020DevelopmentArray}, complementary metal–oxide–semiconductor (CMOS) detectors \cite{Ma2017Photon-number-resolvingGain} and charge coupled device (CCD) detectors. The latter category consists of various schemes to distribute the incoming photons over multiple single-photon detectors \cite{Paul1996PhotonLight, Kok2001DetectionPreparation, Sperling2012TrueDetectors}. Some examples of these schemes are spatial arrays \cite{Eraerds2007, Jiang2007, Divochiy2008}, loop arrays \cite{Banaszek2003, Rehacek2003Multiple-photonDetector, Haderka2004}, and temporal arrays \cite{Achilles2003, Fitch2003, Natarajan2013}.

The spatial array design consists of a symmetric $N$-port coupler that takes the input and ideally distribute the input photons uniformly over the $N$ single-photon detectors. However, the requirements on the number of single-photon detectors grow quickly \cite{Jonsson2020Photon-countingDetectors}, and the limited quantum efficiency of the detectors makes the design difficult to scale. An alternative design that uses fewer resources is the loop-based array \cite{Banaszek2003, Rehacek2003Multiple-photonDetector, Haderka2004}, which only requires one single-photon detector. However, this design results in a non-uniform distribution of photons over the time-bins and most of the light exits in the first few time-bins, which is undesirable.

The temporal multiplexed PNR detector is a device that splits input pulses over time-bins rather than over different detectors. Given that the time-bins are appropriately spaced and that the mean photon number per time-bin is much smaller than unity, it is possible to use single-photon detectors to measure the output. This results in a device that has an output distribution equivalent to a spatial array, but the required number of single-photon detectors is only two. However, by increasing the size of the temporal array, the time required per measurement increase, which reduces the detection rate.

The ability to resolve photon numbers can be quantified in multiple ways, which differ due to the application considered. In this work we mainly consider two different types of PNR capabilities. The first is the ability to find the correct photon number in a single wave-packet, single-shot measurement. This capability is useful in photon number encoded communication or quantum cryptography, where each signal is only sent once. The second is the ability to reconstruct the number distribution of the input source is useful in characterization of low intensity light sources or imaging at low intensities. A detector capable of single-shot PNR measurements is also capable of reconstructing the number distribution, but the former is much more difficult to attain and it is also difficult to quantify it directly.


The work is organized as follows. In Sec. \ref{sec:setup} we present the experimental setup for a $16$ element temporal array with two superconducting nanowire single-photon detectors as detector elements. In Sec. \ref{sec:parameter-estimation} a framework for estimating the mean photon numbers in incident laser light is presented and the limits in resolvable mean photon numbers are predicted using our model. In Sec. \ref{sec:results} the results of the experiments are presented. This includes results from measurements of losses in the multiplexer and in the single-photon detector (in Sec. \ref{sec:detector-losses}), measurements and estimations of PNR capability for multiple input data and for single input data (in Sec. \ref{sec:pnr-capability}), and the bandwidth of the detector (in Sec. \ref{sec:bandwidth}). Finally, in Sec. \ref{sec:summary}, we summarize our findings and draw conclusions.

\section{\label{sec:setup}SETUP}
\begin{figure*}[t]
    \centering
    \includegraphics[width=\linewidth]{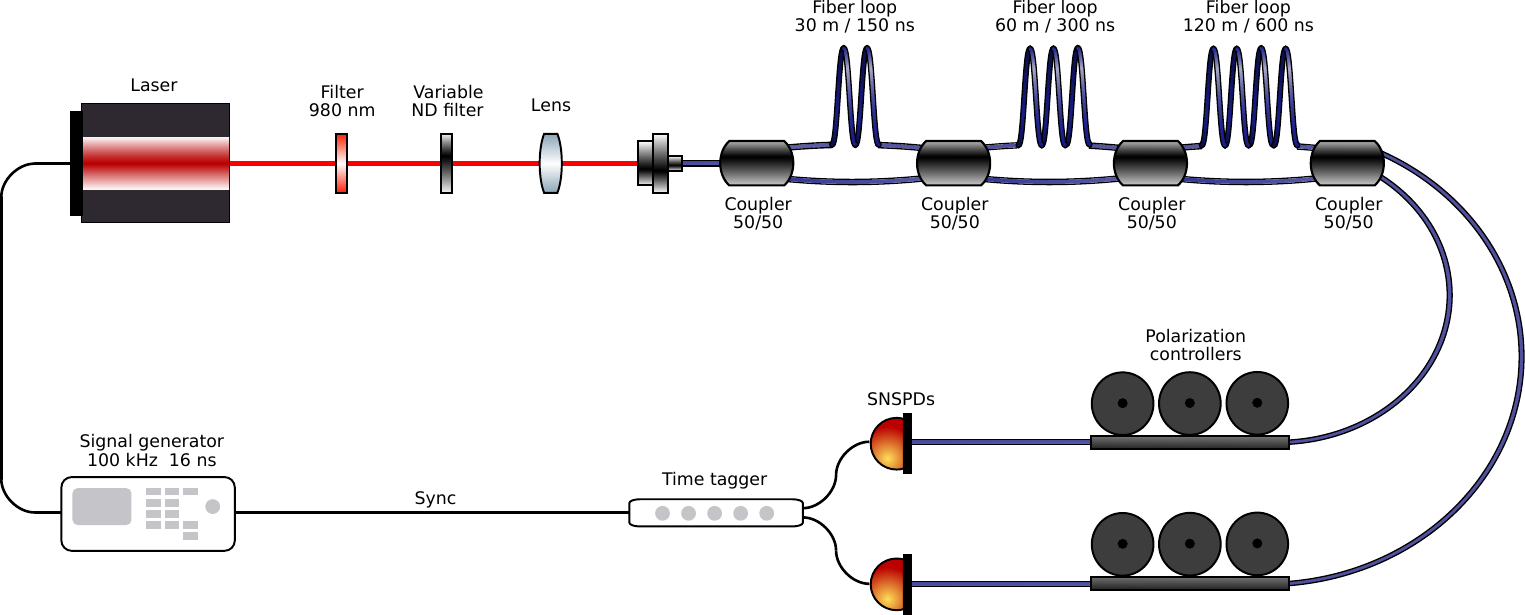}
    \caption{Schematic figure of our setup. The signal generator triggers the laser with a frequency of $\SI{100}{kHZ}$. The laser light is filtered through a $\SI{980}{nm}$ narrow band-pass filter and attenuated with a variable attenuator. The attenuated light is coupled into the multiplexer which splits the light into $8$ time-bins in two output fibers. The polarization of the output is adjusted with polarization controllers and the light is detected with SNSPDs.}
    \label{fig:setup}
\end{figure*}

A schematic image of the setup is presented in Fig. \ref{fig:setup}. A $\SI{980}{nm}$ wavelength semiconductor laser is electrically driven at a repetition rate of $\SI{100}{kHz}$ with a pulse length of $\SI{16.3}{ns}$ using a signal generator. The laser output is filtered through a narrow band-pass filter and subsequently attenuated with a variable, neutral density filter to the few photon regime. These photons are focused with a lens and coupled into a polarization maintaining PANDA fiber PM980-XP. The fiber is connected to a temporal multiplexer made of a series of nominally $50/50$, polarization maintaining fiber couplers, which have been connected with PANDA fiber-loops of lengths between nominally zero to 120 m. 
When splicing the input fiber and the loops to the couplers, great care was taken to align the PANDA fibers ends both transversely and rotationally as to minimize the splicing losses. The multiplexer splits every input laser pulse into $8$ non-overlapping time-bin pulses, which results in an effective detector array size of $16$ elements.

The (linear) polarization state of the output from the multiplexer is rotated using polarization controllers prior to the detectors to maximize the respective count rate from the two superconducting, nanowire single-photon detectors (SNSPD). There are two main reasons for chosing PANDA fibers instead of ordinary single mode (SM) fibers for the multiplexer in spite of the higher losses for the former. The first is that we can consistently make sure that the setup operates in the polarization regime where the SNDPD detection efficiency is the highest. The second is that also fiber couplers typically are polarization sensitive, so that the splitting ratio between the 16 output pulses would depend on the polarization state of the input if SM fibers would have been used. Moreover, mechanical stress and temperature changes may shift the polarization in SM fibers, which makes it difficult to control the polarization.


The SNSPDs are set to operate at a bias current of $\SI{13.00}{\mu A}$, $\SI{10.10}{\mu A}$ and a trigger voltage of $\SI{150}{mV}$, $\SI{50}{mV}$, respectively. The electrical outputs of the SNSPDs are recorded using a quTools quTag time-tagger with picosecond resolution is synchronized with the laser pulses. The outputs from the SNSPDs are time-binned with $\SI{30}{ns}$ windows around the expected arrival times relative to the trigger signal.

\section{\label{sec:parameter-estimation}Parameter estimation}
One potential application for the temporal array detector is to use it to estimate the mean photon number for values where neither single-photon detectors nor power meters are capable of making an accurate measurement. Here we consider the scenario when a Poissonian laser pulse of mean photon number $\mu$ is used as input for the device. The goal is to find an estimator $\hat{\mu}$ and estimate the error $\epsilon = \abs{\hat{\mu} - \mu}$.

We use the maximum likelihood method to find an estimator $\hat{\mu}_{\text{MLE}}$ for the mean photon number $\mu$. To compute this estimator we use that the probability to get $x$ clicks from the SNSPDs after the laser pulse passes through the multiplexer with $n$ effective detector elements is given by \cite{Dauler2009Photon-number-resolutionDetectors, Zhu2019ResolvingDetector, Jonsson2020Photon-countingDetectors}
\begin{equation}
    \Pr(x \mid \mu) = \binom{n}{x} e^{- \mu \eta} (1 - p_d)^n \qty(\frac{e^{\mu \eta / n}}{1 - p_d} - 1)^x,
    \label{eq:poisson-pcd}
\end{equation}
where $\eta$ and $p_d$ are the quantum efficiency and probability for a dark count per time-bin, respectively. The maximum likelihood estimator for the measurement data $\{x_i\}_{i = 0}^{N - 1}$ from $N$ laser pulses is then given by
\begin{equation}
    \hat{\mu}_{\text{MLE}}(\ev{x}) = - \frac{n}{\eta} \ln\qty(\frac{n - \ev{x}}{(1 - p_d) n}),
    \label{eq:mle-poisson-pcd}
\end{equation}
where the expectation value over the experimental data $\ev{x}$ is given by
\begin{equation}
    \ev{x} = \frac{1}{N} \sum_{i = 0}^{N - 1} x_i.
\end{equation}

The error of the maximum likelihood estimator can be divided into two parts, the variance and the resolution error caused by finite number of data points. The variance can be estimated by propagating the sample variance for $\ev{x}$ to $\hat{\mu}_{\text{MLE}}$ and a lower bound can be computed using the Cramer-Rao bound, which gives that
\begin{equation}
    \variance\qty(\hat{\mu}_{\text{MLE}}) \geq \frac{n \qty[(1 - p_d)^{-1} e^{\mu \eta / n} - 1]}{\eta^2 N}.
\end{equation}
Hence, the variance grows at least exponentially with $\mu \eta / n$, which implies that the number of required measurements to reach a certain set variance grows exponentially with the mean photon number and it therefore becomes unfeasible to measure too strong input signals. We also note that an array with size $n$ can measure around $n$ times more photons with the same variance as a single-photon detector.

The resolution error arises from the fact that the quantity $N \ev{x}$ is a non-negative integer less than or equal to $n N$. This implies that the there are in total $nN$ possible finite values $S = \{\hat{\mu}_{\text{MLE}}(0), \hat{\mu}_{\text{MLE}}(1/N), \dots, \hat{\mu}_{\text{MLE}}(n - 1/N)\}$ that the estimator $\hat{\mu}_{\text{MLE}}$ can take and the estimator has a resolution error if mean photon number $\mu \not\in S$. Let us quantify this error by the spacing between the estimator and the next possible value
\begin{equation}
\begin{split}
    \Delta \hat{\mu}_{\text{MLE}}(\ev{x}) &= \hat{\mu}_{\text{MLE}}(\ev{x} + 1/N) - \hat{\mu}_{\text{MLE}}(\ev{x})\\
    &= - \frac{n}{\eta} \ln\qty(\frac{n - \ev{x} - 1 / N}{n - \ev{x}}).
\end{split}
\end{equation}
In the weak light limit when $\ev{x} \ll n$ we notice that the spacing between adjacent estimator values is approximately
\begin{equation}
    \Delta \hat{\mu}_{\text{MLE}}(\ev{x}) \approx \frac{1}{\eta N},
\end{equation}
which is small for sufficiently many measurements. However, for strong light the spacing is not small and the resolution error diverges as $\ev{x} \to n$.

The resolution error and the limited number of outputs effectively limits how large $\mu$ that can be resolved with the detector. The largest value that can be recorded with finite resolution error and a finite estimator is 
\begin{equation}
    \hat{\mu}_{\text{MLE}}(n - 2/N) = \frac{n}{\eta} \ln\qty(\frac{n N (1 - p_d)}{2}).
\end{equation}
For this experiment when $N \approx 10^6$ and the dark counts are negligible we get that the largest resolvable input is $\mu \eta_{\textrm{Max}} \approx 250$ when $n = 16$, while for a (non-multiplexed) single-photon detector $\mu \eta_{\textrm{Max}} \approx 13$.

\section{\label{sec:results}RESULTS}

\subsection{\label{sec:detector-losses}Detector characterization}

\begin{figure}
    \centering
    \includegraphics[width=\linewidth]{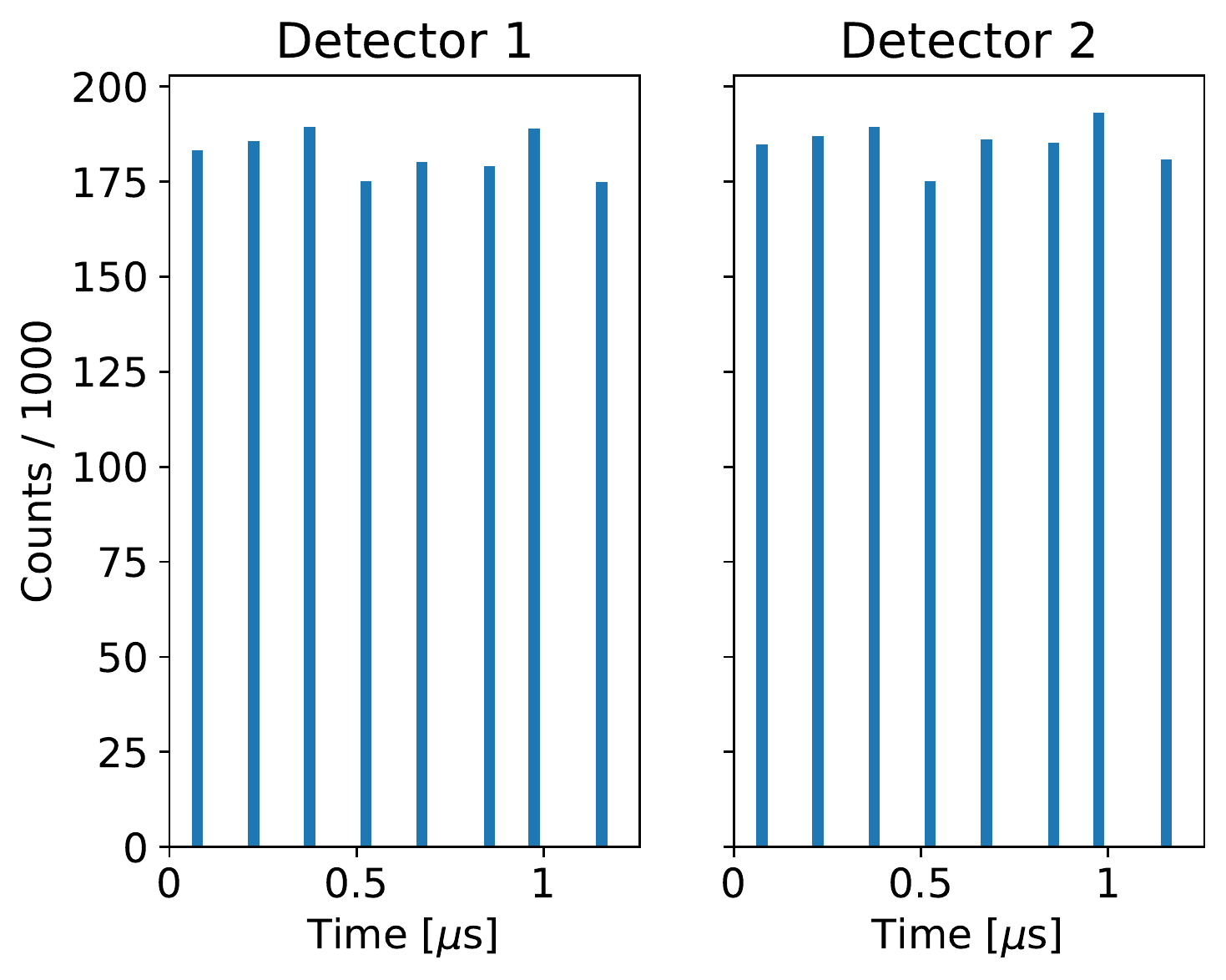}
    \caption{Counts per time-bin is displayed for the detectors. The multiplexer splits the input into $16$, $\SI{30}{ns}$ long, time-bins with close to uniform splitting ratio. Spacing between two following time-bins is approximately of $\SI{150}{ns}$, which sets the requirement that the single-photon detectors must be able to resolve signals of at least $\SI{7}{MHz}$.}
    \label{fig:pulse}
\end{figure}

Losses in the PNR detector can be divided into two types, losses in the temporal multiplexer and limited efficiency of the SNSPD. The former types include fiber losses, losses in splices, and excess loss in the couplers, while losses due to non-unity quantum efficiency belong to the latter type. To estimate the total loss in the multiplexer an optical power meter is used to compare the power at the detectors with the output before the first coupler. These power measurements are conducted without the neutral density filters and the power measurement before the first coupler is done by cutting the fiber. The total loss in the multiplexer is measured to be $\SI{0.63}{dB}$.

The measured loss is well in line with the average expected transmission loss of $\SI{0.66}{dB}$ (the average transmission is 0.86), based on the manufacturers' specifications, where it is assumed that each coupler has a loss of $\SI{0.1}{dB}$ and the fiber has a loss of $\SI{2.5}{dB/km}$ with an average travel distance of $\SI{105}{m}$. Each pulse also passes through 8 fiber splices. However, the splice losses are negligible compared to the other losses.

To measure the quantum efficiencies of the SNSPDs, including the fiber pigtail, the count rate of a laser at $\SI{980}{nm}$ attenuated with a calibrated attenuator is measured. The unattenuated laser power is measured using a power meter and is compared with the measurements conducted with the SNSPDs. The resulting nominal quantum efficiencies are $\SI{50}{\percent}$ and $\SI{64}{\percent}$, respectively, chosen so that the dark count rates are $\SI{0.11}{Hz} \pm \SI{0.04}{Hz}$, $\SI{0.22}{Hz} \pm \SI{0.05}{Hz}$ respectively. Hence, the average, overall quantum efficiency of the whole setup becomes $0.86 \times (0.50 + 0.64)/2 = \SI{49}{\percent}$.

 The probability for a dark count to occur within any of the $\SI{30}{ns}$ time-bins during one input pulse is $\SI{2.6e-9}{}$, $\SI{5.3e-8}{}$, which is negligible for any signal resolvable with the array. Hence it is justified to neglect dark counts in the data analysis.

A histogram displaying recorded detections in one time-bin is presented in Fig. \ref{fig:pulse} and a histogram of all time-bins in one of the output fibers is presented in the inset. The laser pulses have a $\approx \SI{4}{ns}$ FWHM which is about one order of magnitude smaller than the time-bin length. The multiplexer time-bins have an approximate spacing of $\SI{150}{ns}$ which requires the single-photon detectors to have a count rate of at least $\SI{7}{MHz}$ in order not to saturate. The used SNSPDs have a maximal count rate that exceeds $\SI{10}{MHz}$ which suggest that the spacing between the time-bins could be reduced slightly before saturation occurs. However, the SNSPDs experience a quantum efficiency drop when the count rate approaches the maximum count rate. The time-bin spacing was therefore chosen with a margin to guarantee that no extra loss is introduced.

\subsection{\label{sec:pnr-capability}PNR capability}

\begin{figure*}
    \centering
    \subfloat[\label{fig:distribution-od2}]{%
       \includegraphics[width=.33\linewidth]{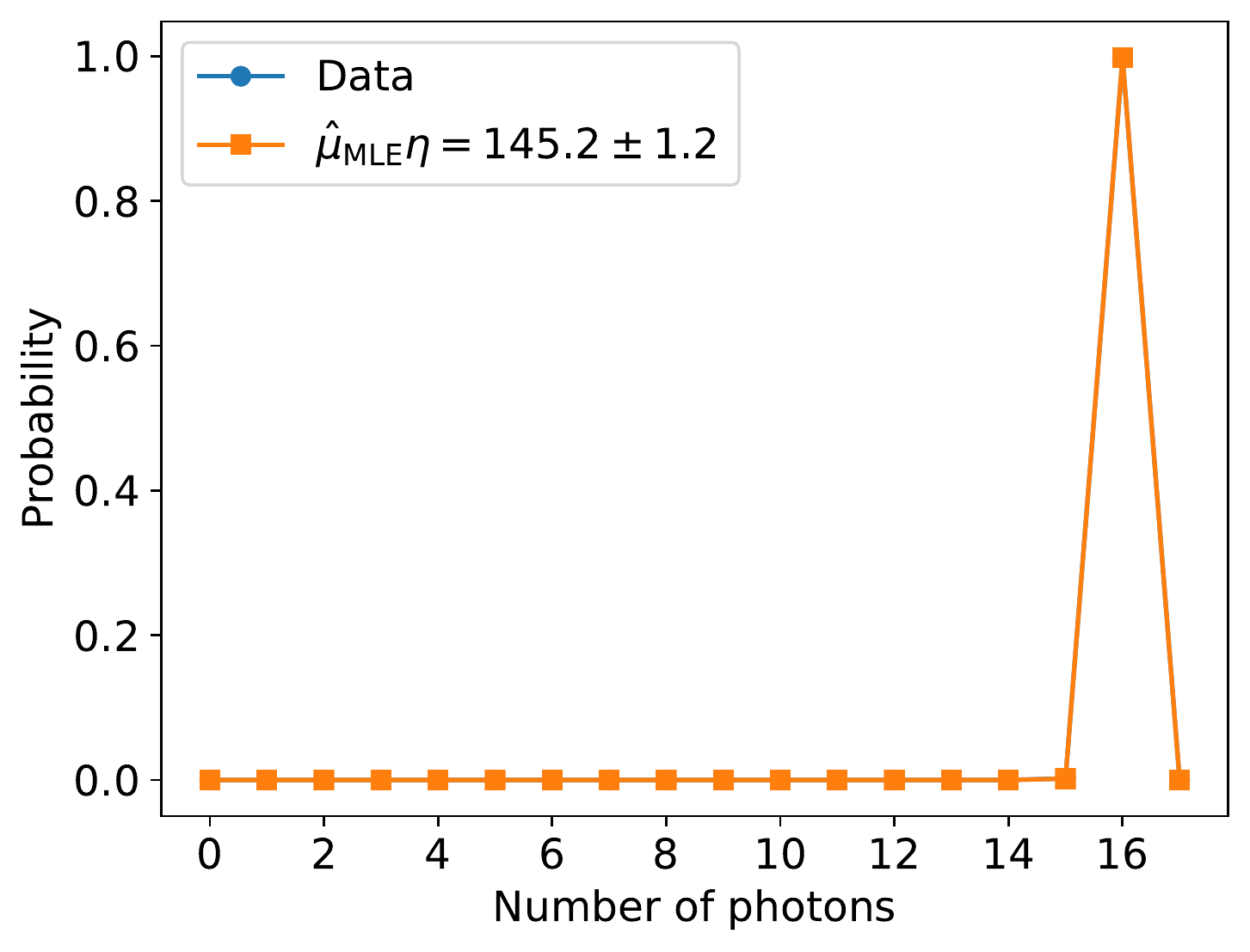}
    }
    \subfloat[\label{fig:distribution-od3}]{%
       \includegraphics[width=.33\linewidth]{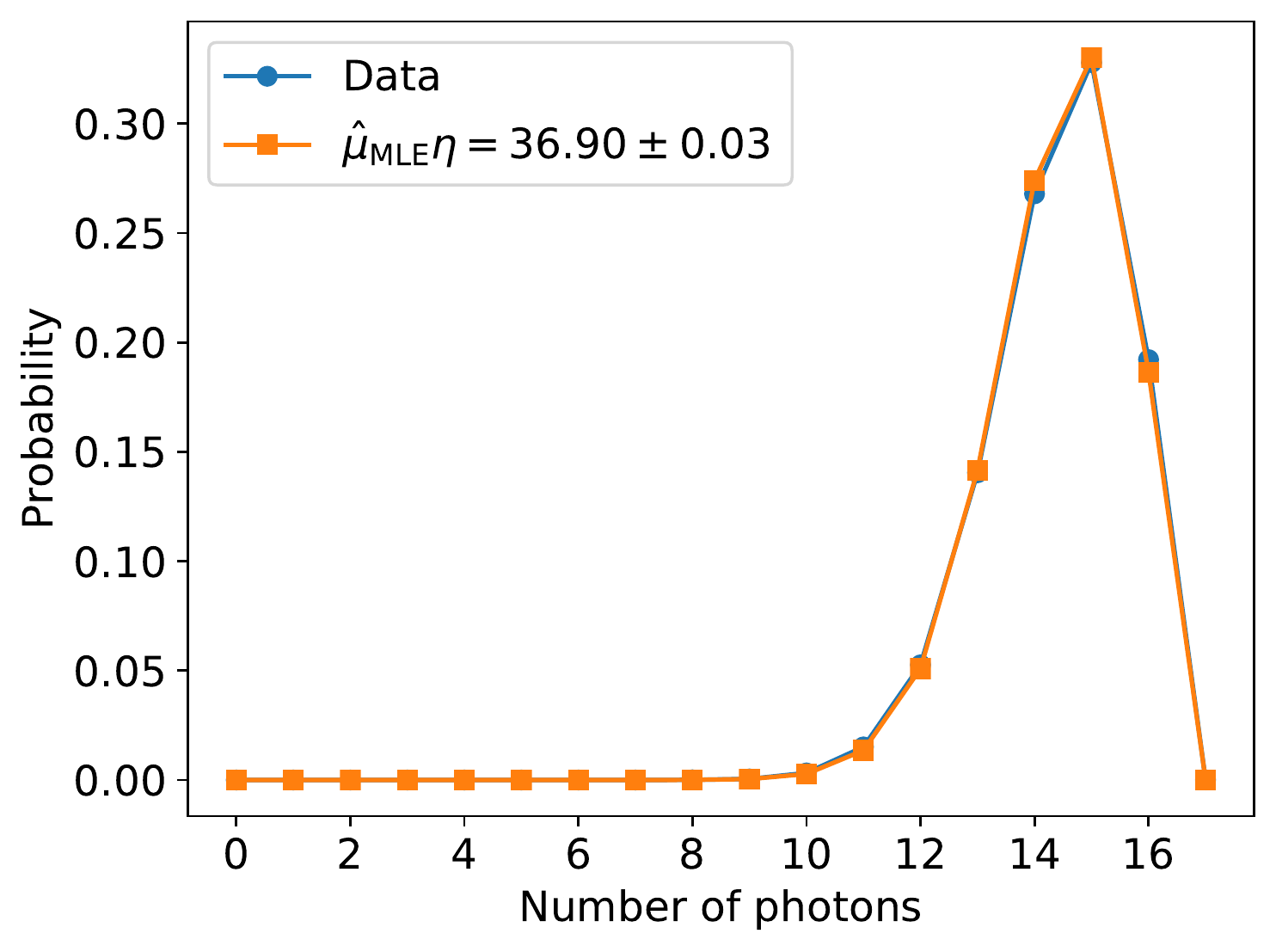}
    }
    \subfloat[\label{fig:distribution-od4}]{%
       \includegraphics[width=.33\linewidth]{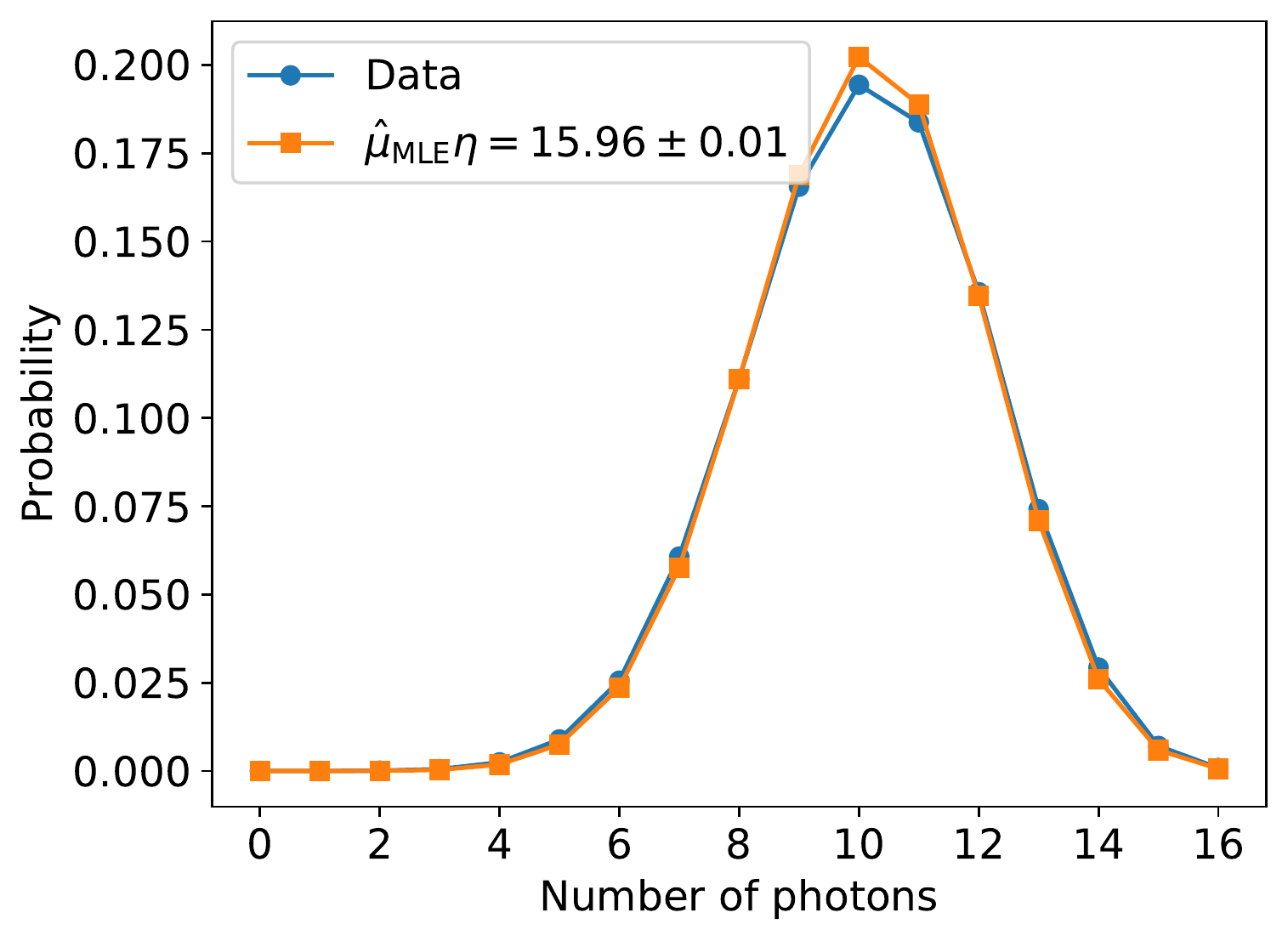}
    }
    
    \subfloat[\label{fig:distribution-od45}]{%
       \includegraphics[width=.33\linewidth]{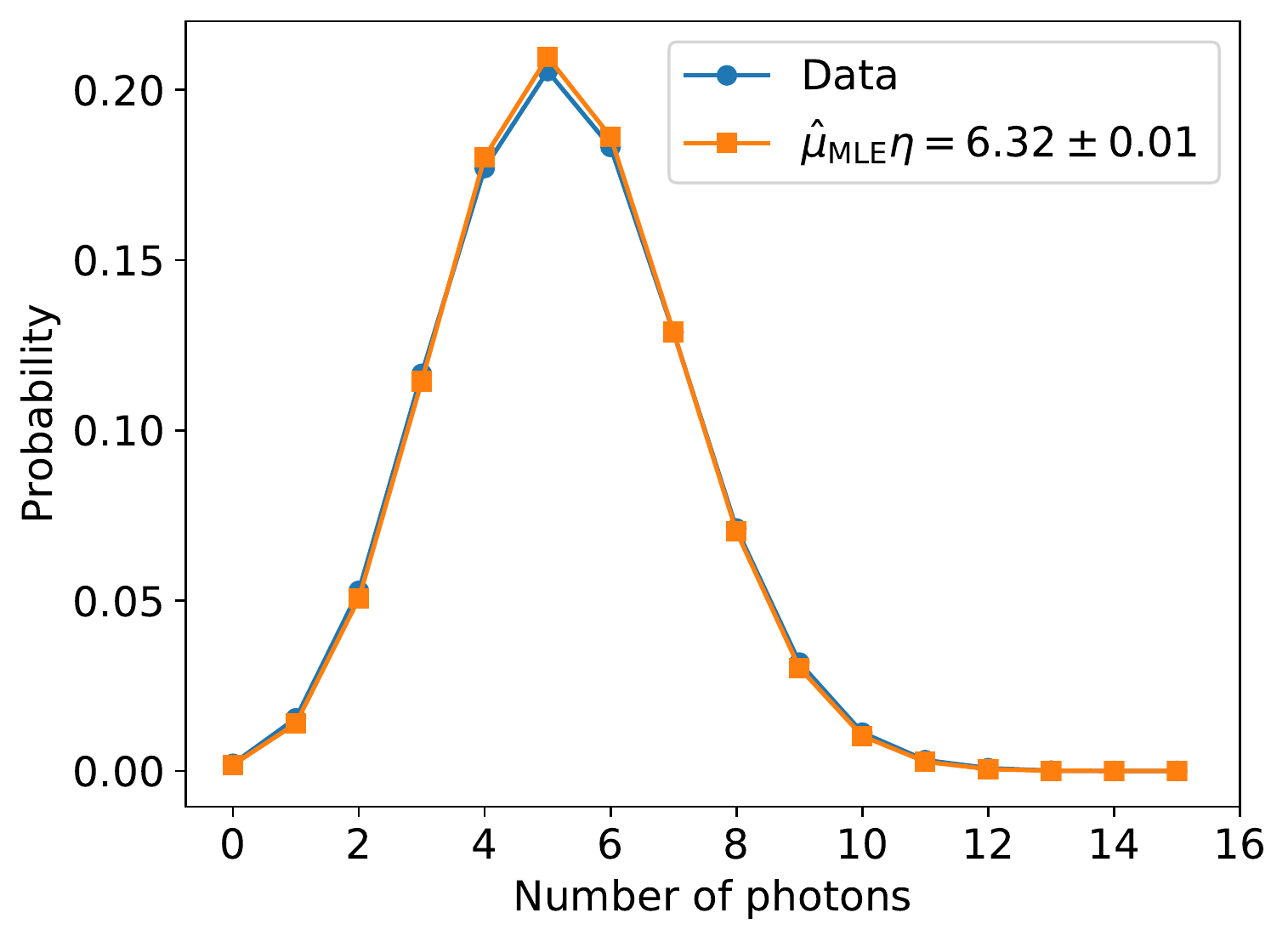}
    }
    \subfloat[\label{fig:distribution-od5}]{%
       \includegraphics[width=.33\linewidth]{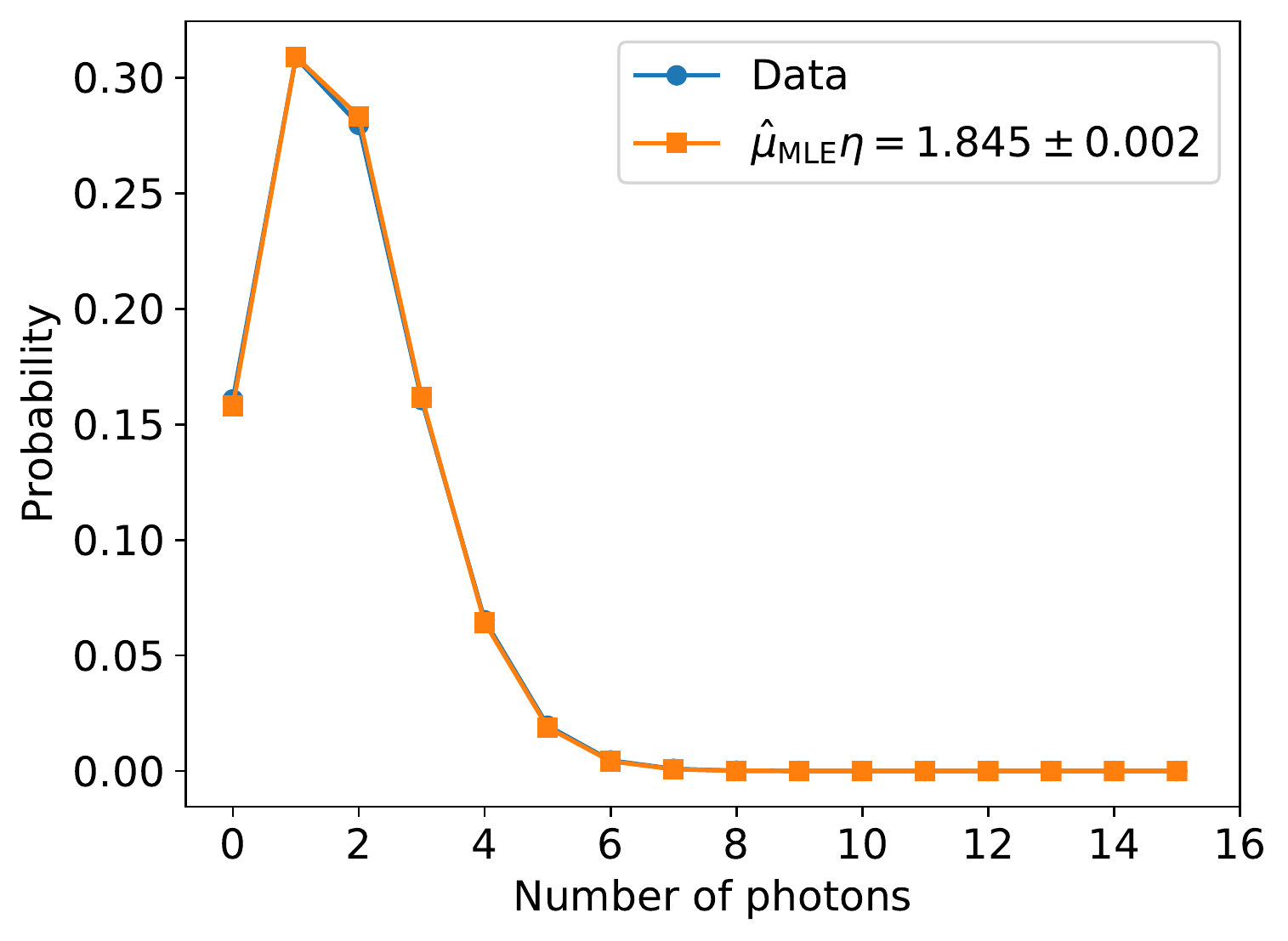}
    }
    \subfloat[\label{fig:distribution-od7}]{%
       \includegraphics[width=.33\linewidth]{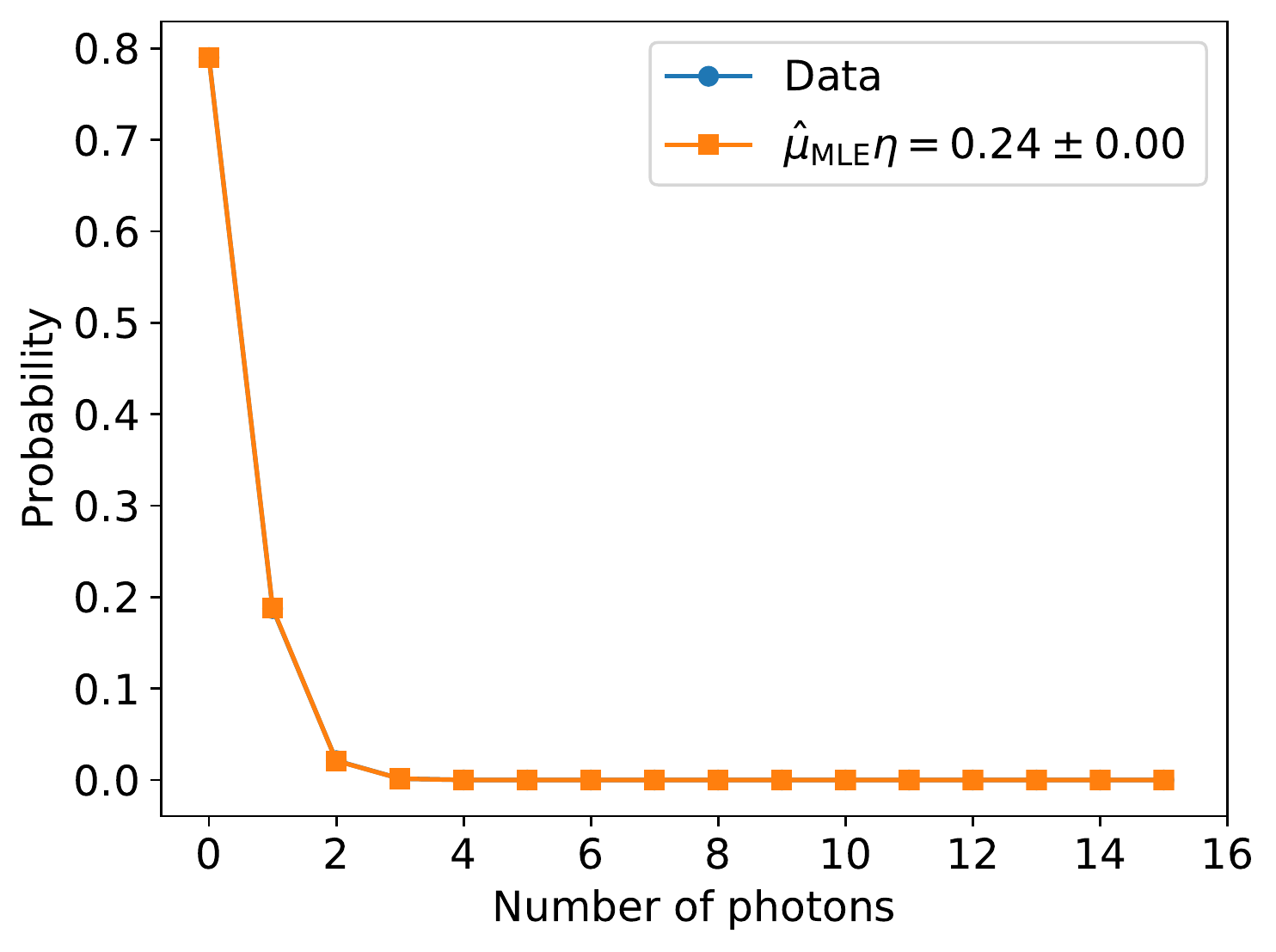}
    }
    \caption{Output count distribution (blue, circles) for different attenuation strengths. Each distribution is fitted to Eq. \eqref{eq:poisson-pcd} using the maximum likelihood estimator under the assumption that dark counts are negligible (orange, squares). The mean photon number estimator is presented in the legends with a one standard deviation estimated uncertainty. The resolution error is at least two orders of magnitude smaller than the standard deviation for these mean photon numbers and is excluded from the presented error. (a) Attenuation is $\text{OD} = 5.0$. The detector is close to saturation and most pulses result in detections in all $16$ time-bins. The existence of pulses resulting in less than $16$ detection events allows for estimation of the mean photon number. The blue curve is completely hidden behind the orange. (b) Attenuation is $\text{OD} = 5.6$. (c) Attenuation is $\text{OD} = 5.9$ (d) Attenuation is $\text{OD} = 6.3$. (e) Attenuation is $\text{OD} = 7.0$ (f) Attenuation is $\text{OD} = 7.9$. For sufficiently small mean photon numbers the distribution approaches a Poisson distribution $\text{Po}(\ev{x})$.}
    \label{fig:distribution}
\end{figure*}

\begin{figure}
    \centering
    \includegraphics[width=\linewidth]{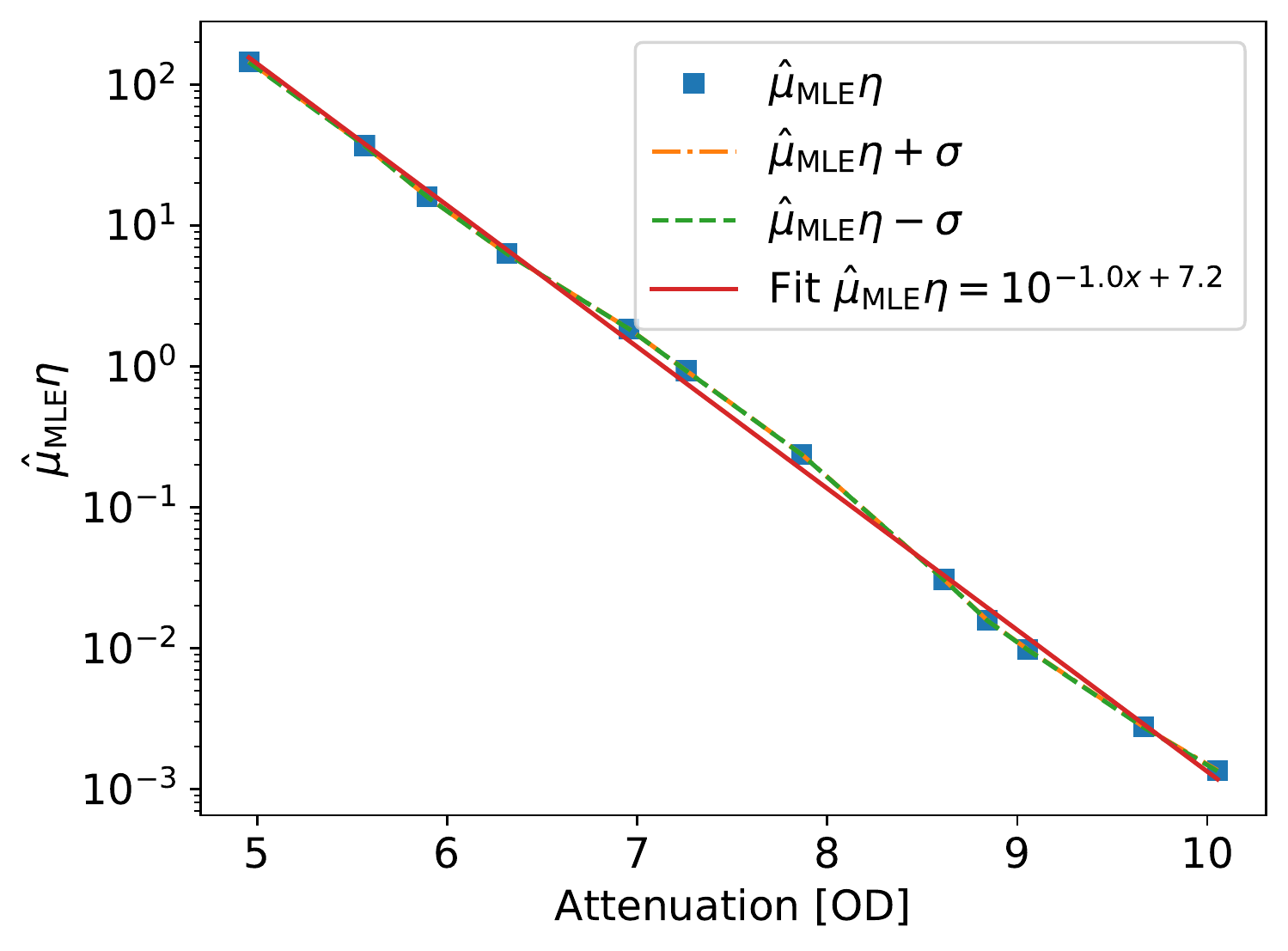}
    \caption{The mean photon-number estimation as a function of the attenuation with an exponential fit. The detector displays good linearity in the measured interval, with a small deviation in the region OD 7 to OD 8. This allows the detector to measure mean photon numbers over five orders of magnitude from $10^{-3}$ to $10^2$.}
    \label{fig:linearity}
\end{figure}

We consider two different types of PNR capabilities in this experiment: The ability to measure the photon number of a single wave-packet and measuring the photon number after many repeated measurements. The former is useful in communication schemes where the data is encoded in photon numbers and the latter is useful when characterizing few photon sources. Here we show the latter experimentally, while the former capability is investigated theoretically using data from the experiments.

The output distribution for different attenuation strengths with corresponding fits are presented in Fig. \ref{fig:distribution}. The fits display good agreement with experimental data suggesting that the array detector setup is well described by equation \eqref{eq:poisson-pcd}. A small deviation between the experimental data and the fit is apparent in Figs. \ref{fig:distribution-od4} and \ref{fig:distribution-od45} which is possibly explained by the fact that the two SNSPDs do not have identical quantum efficiencies as is assumed in the derivation of equation \eqref{eq:poisson-pcd}. The reason for the deviation to be largest for this mean photon number is likely due to the variance of $x$ being maximal for $\mu \eta = n \ln{2} \approx 11.1$ which is close to the value in the experiment. The variance approaches zero for both higher and lower mean photon numbers.

The fitted mean photon numbers as a function of the applied attenuation is presented in Fig. \ref{fig:linearity}. The fit is in good agreement with the experimental data and the fitted exponent is in good agreement with theory. A small deviation from the fit is present in the interval $\text{OD}$ 7 to $\text{OD}8$ which is possibly caused by uncertainty in attenuation or due to instabilities in laser power over time. Overall, the results suggest that the device is capable of accurately estimating the mean, detected photon numbers up to $\hat{\mu}_{\text{MLE}} \eta = 150$, but as predicted in section \ref{sec:parameter-estimation} the upper limit is likely closer to $250$. The estimated number can then be used to find the absolute mean photon numbers by carefully calibrating the quantum efficiency of the detectors.

The experimental setup is not very suitable to investigate the multiplexed detector's ability to resolve photon numbers in a single wave-packet. In order to do this, the laser should be replaced by a deterministic few photon source. However, the present setup makes it is possible to estimate the probability for a correct classification using the detector parameters. We assume that the detector is well described by \eqref{eq:poisson-pcd} which is supported by the preceding results in this subsection. It then holds that the probability to get $x$ clicks when the input is a Fock state of $m$ photons is given by \cite{Fitch2003, Jonsson2020Photon-countingDetectors}
\begin{equation}
\begin{split}
    \Pr(x \mid m) &= \frac{1}{n^m} \binom{n}{x} \sum_{l = 0}^x (-1)^l (1 - p_d)^{n - x + l} \binom{x}{l}\\
    &\times [n - (n - x + l)\eta]^m. 
\end{split}
\end{equation}
For this setup, the total loss (loss due to non-unity quantum efficiency combined with the loss in the multiplexer) is $\eta = 0.49$ and $p_d = 0$ since dark counts can be neglected. Given these parameters it is only possible to resolve almost one photon with at least $\SI{50}{\percent}$ probability for a correct classification. This shows, that in spite of a rather elaborate choice of components and a working wavelength on the short side of the near-infrared wavelength range, it is still difficult to achieve true, single-shot PNR performance. Increasing the SNSPDs' quantum efficiencies to $\eta > 0.9$ would still only allow single-shot PNR performance with up to two photons using this 16-array multiplexer.


Next, we analyze the upper limits of our multiplexer to its PNR performance by assuming idealized components. First we consider the scenario when the single-photon detectors are ideal and have unit quantum efficiency. The effective quantum efficiency for our setup in this scenario is $\eta = 0.86$, all due to the loss in the multiplexer. In this case it is possible to measure up to $3$ photons with more than $\SI{50}{\percent}$ probability for a correct classification for the photon numbers 0 to 3 \cite{Jonsson2020Photon-countingDetectors}.

Finally, we consider the (unrealistic) scenario when both the single-photon detectors and the multiplexer are ideal with no losses. The effective quantum efficiency is in this scenario $\eta = 1$ and the detector is only limited by the probability that more than one photon ends up in the same time-bin. In this scenario it is possible to resolve up $5$ photons with $> \SI{50}{\percent}$ success \cite{Jonsson2020Photon-countingDetectors}. This demonstrates the difficulty to resolve photon numbers in a single-shot measurement scenario.

\subsection{\label{sec:bandwidth}Bandwidth}
\begin{figure*}[t]
    \centering
    \includegraphics[width=\linewidth]{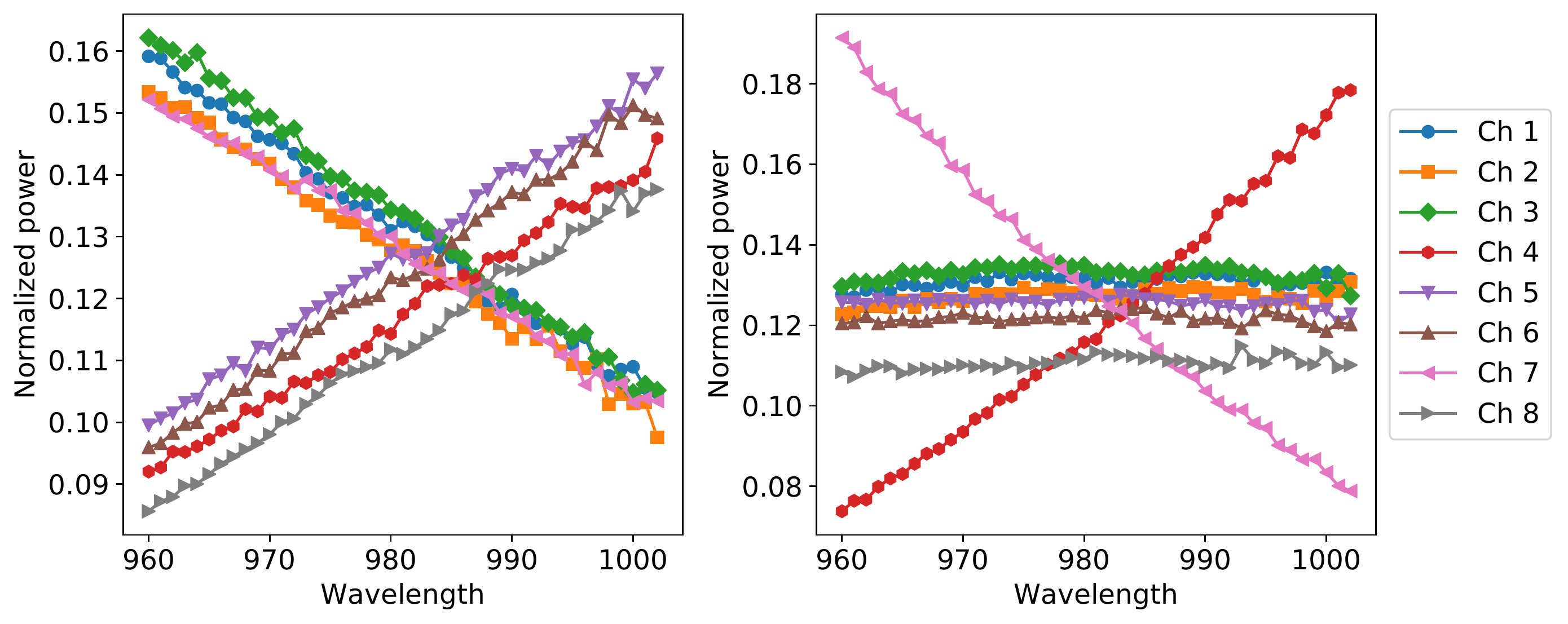}
    \caption{Power in the 8 time-bins for each single photon detector normalized against the total power as a function of the wavelength. Both detectors show a linear relationship between normalized power and wavelength. (Left) The normalized power in detector 1. (Right) The normalized power in detector 2.}
    \label{fig:bandwidth-detector}
\end{figure*}

The single-photon detectors used to build the temporal array are wavelength dependent and the temporal array is therefore also wavelength dependent. In addition, a temporal-array detector built using fiber couplers also experiences a wavelength dependence due to the wavelength dependence of the coupling ratio. The latter dependence effectively reduces the array size by changing the splitting ratio for the couplers. This in turn, reduces the dynamic range and PNR capability of the array detector.

In Fig. \ref{fig:bandwidth-detector} the normalized power in each time-bin is presented as a function of the wavelength.  The figure shows that commercial fiber couplers still exhibit quite a bit of individual behavior. The measurements show that the relation between the coupling ratio and the wavelength is linear, at least within a few percent's deviation from the design wavelength. Given the linear relationship, the transmission from the lower input port to the lower output port or the upper input port to the upper output port can be modelled as $T_{ll} = T_{uu} = 1/2 + a \Delta \lambda$. Here $a$ is a constant determining how fast the transmission changes per unit wavelength, $\Delta \lambda$ is the offset from the operational wavelength and by assumption $\abs{a \Delta \lambda} \ll 1/2$ in the interval considered. Unitarity of the coupler implies that the transmission from lower to upper or vice versa is given by $T_{ul} = T_{lu} = 1/2 - a \Delta \lambda$.

The average power of the different time-bins can be computed using the transmissions and the total input power. To linear order in $a \Delta \lambda$ we get that the normalized power in the time-bins in the first detector are given by
\begin{equation}
    \frac{P_{1, \pm}}{P_{\text{tot}}} = \frac{1}{16} \pm \frac{1}{4} a \Delta \lambda + \order{(a \Delta \lambda)^2}
\end{equation}
where half of the time-bins correspond to $P_{1, +}$ and the other half to $P_{1, -}$. The normalized powers in the second detector are given by
\begin{align}
    \frac{P_{2, 0}}{P_{\text{tot}}} &= \frac{1}{16} + \order{(a \Delta \lambda)^2},\\
    \frac{P_{2, \pm}}{P_{\text{tot}}} &= \frac{1}{16} \pm \frac{1}{2} a \Delta \lambda + \order{(a \Delta \lambda)^2},
\end{align}
where six time-bins correspond to $P_{2, 0}$ and the $P_{2, +}$, $P_{2, -}$ have one time-bin each. The non-equal behavior in the first and second detector are due to the symmetry being broken when one input port of the first 50/50 coupler is chosen over the other. If equal power is incident in both input ports in the temporal multiplexer (but in a manner so that no interference between different pulses ensue) the output is also expected to be symmetric.

In the wavelength interval considered the array size of our detector effectively decreases to $11$ elements as $\abs{\Delta \lambda}$ grows sufficiently large. These $11$ elements sustain the same power or increase in power as the wavelength changes and they can therefore still be used to measure the input signal. The remaining $5$ time-bins have a power that is quickly reduced with wavelength and their contribution to the measurements therefore decrease as the wavelength differs sufficiently from $\SI{980}{nm}$.

\section{\label{sec:summary}SUMMARY}

In this paper we present an experimental realisation of a temporal detector array with $16$ time bins, and SNSPDs as detectors. We investigate the theoretical PNR capabilities of this detector and determine an upper bound for the mean photon number that can be resolved by the detector when multiple measurements are allowed. We conclude that our detector is able to resolve one order of magnitude larger mean photon numbers than the SNSPD detector alone is able to do with the same number of measurements. In numbers, our detector can measure the average photon number in pulses between $10^{-3}$ to $10^{2}$ photons if data is collected from a sample of around $10^6$ identical pulses. The precision improves with the sample size, while the accuracy is mainly determined by how well the quantum efficiency of the used SNSPDs can be determined. In our case the imprecision after $10^6$ samples is estimated to be $\SI{1}{\percent}$, while the accuracy is difficult to estimate since there are many factors that add uncertainty to the overall loss estimations.

Our experimental investigation of the PNR capabilities shows that the agreement with the theoretical model is excellent and that the detector is linear over $5$ orders of magnitude when estimating the mean photon number over multiple measurements. This suggests that the theoretical predictions of the maximal number of photons is reasonable. Furthermore, error estimations show that the predictions made with the detector have low variance and a statistical bias caused by discretized readout is negligible. 

While we are using a pulsed input, the detector could in principle be used to measure the power in a (slowly varying) cw signal. To do this, one would use a periodic gating and use the gating frequency, the detector on-off ratio, and the overall quantum efficiency to translate the photo counts to optical power at a given wavelength. Used this way, our detector could measure powers in the range 0.1 pW to 10 nW. However, by increasing the gating frequency or the on/off-ratio (without coming close to unity), even lower powers could be detectable. The gating frequency sets a upper limit to how rapid changes in the signal can be detected. The detector's dark count rate sets the fundamental lower limit to the detectable optical power. At a dark count rate of 1 Hz and a SNR of unity the detection limit becomes around $10^{-19}$ W. However, such extremely weak signals carry non-negligible intrinsic fluctuations (due to shot-noise), so from a practical viewpoint one would probably not want to operate the detector below $10^{-17}$ W.

On the negative side, in spite of substantial efforts to maximize the overall quantum efficiency of the detector, our detector is not capable of single-shot photon-number resolution of pulses since its average, overall efficiency is $0.49$. To distinguish between zero, one, and two photons with a minimum success rate of 0.5, the quantum efficiency of the detector needs to be $\geq 0.85$. At telecom wavelengths fiber propagation losses are lower than at $\SI{980}{nm}$. However, this lower loss is offset by a decreasing quantum efficiency of the detectors as the wavelength increases \cite{Natarajan2013}.

We also investigate the bandwidth of the detector and show that the time-bins have different sensitivities for changes in wavelength. This effect is due to the couplers' changing splitting ratios when the wavelength changes and some time-bins correspond to paths where this effect accumulate, while other time-bins correspond to paths where the effect cancel. This shows that the power in $11$ of the $16$ time-bins is unchanged or increases as the wavelength changes with $\SI{20}{nm}$ around $\SI{980}{nm}$.

To improve array-based PNR detectors further, one would need to use detectors with very high quantum efficiency. It would also help to space the time-bins denser, resulting in shorter loops with lower propagation loss. However, the length of the loops in the multiplexer needs to be long enough to give the single-photon detectors time to recover without reducing the quantum efficiency. Ideally, one would like to make the whole device monolitically integrated as a photonic circuit. However, for this to happen, the detector recovery-time needs to be shortened to the sub-ns range in addition to solving the problem of integration between photonic and superconducting circuits.

\section*{\label{sec:acknowledgments}Acknowledgments}
M.J. and G.B. acknowledges funding from Knut and Alice Wallenberg Foundation through the grant Quantum Sensing and from the Swedish Research Council (VR) through grant 621-2014-5410.

M.S. acknowledges funding from the Swedish Research Council (VR) through grant 2018-03457.

S.G. and V.Z. acknowledges support of the ATTRACT project funded by the EC under Grant Agreement 777222, funding from the Knut and Alice Wallenberg Foundation Grant “Quantum Sensors” and support from the Swedish Research Council (VR) through the VR Grant for International Recruitment of Leading Researchers (ref. 2013-7152) and Research Environment Grant (ref. 2016-06122).

Figure \ref{fig:setup} was created with ComponentLibrary \cite{componentlibrary}.

\appendix

\end{document}